**THE DIRAC FIELD AT THE FUTURE CONFORMAL SINGULARITY**

Michael Ibison

Institute for Advanced Studies at Austin, USA

### 1. INTRODUCTION

In the flat space Friedmann-Robertson-Walker (FRW) Cosmology expressed in the coordinate system with line element $ds^2 = d\tau^2 - a^2(\tau)d\mathbf{x}^2$, $\tau$ is the proper time of a 'fundamental observer' equipped with a suitably-defined 'laboratory clock' at rest with respect to the frame established by the Cosmic Microwave Background. Measured according to this time The Universe will expand forever, asymptotically approaching an exponential growth $a(\tau) \to \exp(H\tau)$, where $H$ is the Hubble parameter. Descriptions and explanations of Cosmology commonly adopt this time implicitly - though GR is coordinate independent of course. For example the cosmological (as opposed to Doppler) red-shift of distant galaxies is usually described as due to the effect of the expansion of space during the time light – or a photon – is in flight from the distant star on its way to Earth. Unless properly qualified such explanations give the false impression that a physical statement is being made about the effect of expansion on light, without recognizing that the explanation makes use of a projection of the underlying physics onto a particular coordinate system. In this instance the physical – coordinate independent – essence of the phenomenon depends on the differential evolution of electron mass and photon energy, and not on the photon energy by itself. So for example a perfectly good alternative explanation for the same observation is that the Lyman Alpha lines of local hydrogen, i.e. at the time of reception, are more separated energetically (bluer) then those at the time of emission, with light having suffered no red-shift at all whilst in flight. This alternative picture comes from projecting the physical process onto the conformal coordinate system having line element $ds^2 = a^2(t)(dt^2 - d\mathbf{x}^2)$. In contrast with the 'traditional' coordinate system, in the conformal system $(-g)^{1/2}$ times the energy density of matter increases in proportion to the scale, giving rise, effectively, to an increase in the rest mass of a free electron. Of course there are infinitely many other coordinate systems. But this one example serves to illustrate the difference between a process and its projection.

One does not expect to find any new physics simply by changing the coordinate system; a coordinate transformation in GR is analogous to a coordinate rotation in Euclidean geometry. Yet there is a concern. The FRW system has no future boundary, whereas the conformal scale factor is singular at a finite conformal time; the exponential expansion written conformally has asymptote $a(t) \to (1 - Ht)^{-1}$. One might wonder if perhaps this is a coordinate singularity with no physical consequence and can be ignored? As implied above (though with some qualifications) EM radiation is unaffected by the expansion expressed in the conformal system – perhaps it is unaffected also by the singularity? And how is fermionic matter affected by the singularity?

Penrose, and Friedrich (Friedrich, H., 2002; Penrose, R., 1963) have considered in detail the consequences of the conformal singularity for gravitational radiation in particular. The focus here is on the consequences for the *Dirac field*. We will show below that Dirac's equation in conformal spacetime with conformal scale factor $f(x)$ can be written

$$\left(\gamma^\mu\left(-i\partial_\mu + eA_\mu(x)\right) + f(x)m\right)\psi(x) = 0 \,. \qquad (1)$$

Of interest is the manner in which $f(x)$ introduces a dependency on absolute coordinates and so changes the way the discrete symmetry operations affect the equation as a whole. A conformal factor due to cosmological expansion has a different status than the vector potential because it is a fixed background affecting all particles everywhere. Arguably it should be considered as hard-wired into the Dirac equation in the same manner that the (constant) mass term is ordinarily considered a fixed feature. In fact since it multiplies the mass one can take the position that a *coordinate dependent mass* is a universal property of the Dirac equation. Obviously this point of view pertains specifically to the equation expressed in conformal coordinates.

We are interested in the ways the conformal factor influences behavior: 1) On the local modification of discrete symmetries, 2) On the relationship between pre and post-singularity wavefunctions, 3) The boundary conditions on the wavefunction and / or the topologies of Cosmological spacetime required in order that the wavefunction behave nicely *through* the singularity. These topics are covered in the subsequent sections as follows. Section 2 reviews the symmetry-breaking effects of the conformal metric on the discrete symmetries normally present in Minkowski spacetime. The latter are reviewed in Appendix A. Section 3 looks at the behavior of the conformal scale factor near and through the singularity as determined by the Friedmann equation for a conformally-expressed metric in the flat space Robertson-Walker spacetime. Section 4 gives a very brief review of EM in conformal spacetime. The affect of the conformal metric on the Dirac equation in general, and the wavefunction in particular are covered in Sections 5 and 6 respectively. Those findings are then applied in Section 7 to the particular case of a conformal representation of the de Sitter spacetime. Those results are further specialized in Section 8 in an analysis of the wavefunction near and through the singularity. (With some qualifications, all vacuum-dominated Robertson-Walker spacetimes asymptote to the de Sitter evolution). Section 9 outlines the alternatives for peaceful coexistence between the Friedmann equation and the Dirac wavefunction under the presumption that the post-singularity universe is not a redundant copy of the pre-singularity universe.

## 2. INVERSION SYMMETRIES

### 2.1 Systematic Symmetries

An inversion operation may be a symmetry of the whole of *system* of physical interactions if applied universally. By universally we mean here not just over all space and time, but to all particles. It is easy to see for example that the system of QED must be invariant under charge conjugation. QED is invariant also under parity and time reversals, independently. Since the system as a whole is invariant, it follows that an inversion applied universally to a physically legal universe of particles and their interactions generates another universe, legal under the rules of QED. It is different matter however to determine if this implies a symmetry (point symmetry or local symmetry) of the Dirac wavefunction, either free, or in the presence of an interaction.

### 2.2 Point Symmetry and Local Symmetry in The Dirac Equation

A free particle solution in Minkowski spacetime $\psi(t,\mathbf{x})$ i.e. obeying (1) with $A_\mu(x) = 0$, $f(x) = 1$,

$$\left(-i\gamma^\mu \partial_\mu + m\right)\psi(x) = 0 \qquad (2)$$

turns out to have symmetric partners associated with time reversal and parity inversion. Having fixed a coordinate system with a particular origin, technically, the replacement $\mathbf{x} \to \mathbf{x}' = -\mathbf{x}$ has the specific meaning of inverting the coordinates through the spatial origin $\mathbf{x} = \mathbf{0}$. Therefore the presence of a symmetry under this operation implies that for every solution $\psi(t,\mathbf{x})$ there exists another solution $\psi'(t,\mathbf{x})$ of the same equation that is somehow related to the solution at $\psi(t,-\mathbf{x})$. For example the relationship may be of the form $\psi'(t,\mathbf{x}) = U\psi(t,-\mathbf{x})$, where $U$ is some fixed 4x4 matrix. If the origin remains where it was before the inversion this relates solutions that are spatially separated, depending on their distance from the origin. That is, $\psi'(t,\mathbf{x}) = U\psi(t,-\mathbf{x})$ is a *point symmetry*. To be concrete, if the origin were at the center of the Milky way, then the symmetry implies that for every Earth-based particle with wavefunction $\psi(t,\mathbf{x})$ there may be another particle with wavefunction $U\psi(t,-\mathbf{x})$ located on the 'other side' of the Milky Way at a distance of about 100,000 light years from Earth. Usually though this is not what is meant by parity inversion symmetry in the context of the Dirac equation. Instead it is understood that there is a freedom to combine the replacement $\mathbf{x} \to \mathbf{x}' = -\mathbf{x}$ with an arbitrary translation to bring the symmetric partner to the same location as the original. This is possible because (2) is translation invariant. In practice this freedom is used to move the origin to the location of the original particle before applying the operation $\mathbf{x} \to \mathbf{x}' = -\mathbf{x}$, with the outcome that inversion can be treated as an entirely local operation. The combination of translation invariance with parity inversion invariance converts the point symmetry to a *local symmetry*.[1] When the vector potential is absent one usually expects to see the local version of the time and parity symmetries. Generally, the effect of a (universal) conformal scale factor is to destroy the translation invariance to some degree in the same manner as would the presence of a vector potential. With translational invariance gone only the point symmetry versions remain, these now to be interpreted with respect to the absolute coordinate system established by the metric and, in particular, the conformal singularity. Consequently in the following we will be interested in how the conformal factor affects the local symmetry, and additionally in the consequences of its replacement with a point symmetry.

---

[1] This point of view is applicable only to a single particle, which is our sole interest here.

## 3. THE FRIEDMANN EQUATION IN CONFORMAL SPACETIME

### 3.1 Conformal Forms of the Scale Factor

We consider in parallel the two forms of Cosmological scale factor

$$f(x) \in \left\{ a(t), a(t/x^2)/x^2 \right\} \tag{3}$$

where $a$ is an arbitrary function common to both cases determined from solution of the Friedmann equation. The notations used here are $x = \{x^\mu\} = (t,\mathbf{x})$, $r = |\mathbf{x}|$. We note in passing that the first of these admits the alternative form

$$f(x) = \frac{1}{t} b\left(\frac{t}{x^2}\right) \tag{4}$$

where $b(z) = za(z)$. Each of (3) is associated with a line-element

$$ds^2 = f^2(x) dx^2 \tag{5}$$

which is why they are called conformal. The coordinates used for these definitions are not the same though we have used the same symbols - they are related by the coordinate transformation

$$t \to t/x^2, \quad r \to r/x^2 \tag{6}$$

with the angle variables left unchanged. Note that the transformation is symmetric in that it can be applied to take the first of (3) to the second and vice-versa. Because a coordinate transformation exists between these two forms they are equivalent from a GR point of view, though they may *imply* different topologies. When expressed as a Robertson-Walker spacetime the second of these has zero spatial curvature. That is, it can be written in the form

$$ds^2 = d\tau^2 - a'^2(\tau) d\mathbf{x}^2 \tag{7}$$

where $d\tau = a(t)dt$ and $a'(\tau) = a(t)$ is a (new) arbitrary function. Since the two systems are related by a transformation the first of (3) can also be put into the form (7) and therefore also has zero spatial curvature in that context. (Spatial curvature is not a coordinate-independent quality of a metric. It is sufficient to note that all Robertson-Walker spacetimes can be expressed conformally, removing therefore the spatial curvature from the $K = \pm 1$ spacetimes - see (Ibison, M., 2007).) In the form $f(x) = a(t)$ the (hyper) surfaces of cosmological simultaneity are the hyper-planes $t$ = constant. In the case $f(x) = a(t/x^2)/x^2$ then $t = kx^2$ for some constant $k$ and then

$$\left(t - \frac{1}{2k}\right)^2 - r^2 = \left(\frac{1}{2k}\right)^2. \tag{8}$$

Hence the surfaces of simultaneity are paraboloids, including the 'final surface' that is the conformal singularity. Despite superficial appearances therefore, both forms in (3) admit three independent translational isometries.

### 3.2 Two Branches

The evolution of the scale factor is decided by the Friedmann equation plus equations of state for the various contributions.[2] In conformal coordinates this is

$$\frac{1}{H^2}\left(\frac{da}{dt}\right)^2 = \Omega_{EM} + \Omega_m a + \Omega_\Lambda a^4. \tag{9}$$

Each of the $\Omega$ is an energy density normalized so that their sum is unity; $\Lambda$ denotes the vacuum contribution. Present estimates are (Nakamura, K. & et al, 2010)

---

[2] Either the equations of state or the second Friedmann equation involving the pressure.

$$H^{-1} = 9.78 / .72 = 13.6 \quad \text{Gyr}$$
$$\Omega_\Lambda = 0.74, \quad \Omega_m = 0.256, \quad \Omega_{EM} = 4.76 \times 10^{-5} \ . \tag{10}$$

Let us write the Friedmann equation as

$$\frac{1}{H^2}\left(\frac{da}{dt}\right)^2 = f(a) \tag{11}$$

where $f(a)$ is a dimensionless function of the scale factor. Upon integration one has

$$t_+(a) = \frac{1}{H}\int^a \frac{da'}{\left|\sqrt{f(a')}\right|} \quad \text{and} \quad t_-(a) = -\frac{1}{H}\int^a \frac{da'}{\left|\sqrt{f(a')}\right|} \tag{12}$$

I.E. $t$ has two (single-valued) branches $t_+(a)$ and $t_-(a)$ which are true functions. (The square root operation is discussed in more detail below.) Since the integrand is always positive or zero the $t_\pm(a)$ are both monotonic with $a$. Therefore each function is invertible, and in each of these the scale factor can be expressed as a function of time. Let us write the Taylor-Laurent series expansion of $f(a)$ about $a = 0$ as

$$f(a) = c_m a^m + c_{m+1} a^{m+1} + \ldots + c_{n-1} a^{n-1} + c_n a^n; \quad n > m \tag{13}$$

and set $a = 1$ at the present – finite – time. Then the integrals in (12) converge to a finite future $t$ as $a \to \infty$ provided $n > 2$. In that case there is a singularity in the scale factor in finite conformal time. Similarly the integrals converge to a finite $t$ as $a \to 0_+$ provided $m \geq 0$. In that case the Big Bang occurred at a finite conformal time in the past. In (9) $m = 2$, $n = 4$, and both of these conditions are met; the universe has a finite conformal duration. Defining

$$\phi(a) := \frac{1}{H}\int_\infty^a \frac{da'}{\left|\sqrt{f(a')}\right|} \tag{14}$$

the solutions (12) can be written

$$t_+(a) = \phi(a) + t_+(\infty) \quad \text{and} \quad t_-(a) = -\phi(a) + t_-(\infty) \tag{15}$$

where we have chosen to set whatever initial conditions we intend to apply at the conformal singularity (which will now be presumed to exist). Let us set the clocks to zero there - $t_+(\infty) = t_-(\infty) = 0$ - rather than at the Big Bang. With this, and inverting (15), the inversion of the two branches gives

$$a_+(t) = \phi^{-1}(t) \quad \text{and} \quad a_-(t) = \phi^{-1}(-t) \ . \tag{16}$$

### 3.3 Parity of the Scale Factor

Usually there would be no motivation to entertain solutions of the Friedmann equation for negative values of the scale factor. But since the conformal boundary occurs at a finite time and the scale changes sign thereafter, this would appear to be an oversight. We will see below that the energy of Dirac matter changes sign through the boundary, with the result that the product $\Omega_m a(t)$ is always positive. Consequently (9) is more accurately written

$$\frac{1}{H^2}\left(\frac{da}{dt}\right)^2 = \Omega_{EM} + |\Omega_m a| + \Omega_\Lambda a^4 \tag{17}$$

Now the right hand side has even parity. Consequently $f^{1/2}(a) = f^{1/2}(-a)$ so that the even parity survives the square root operation. No questions arise as to the meaning of the square root operation here and in (14) because $f(a)^{1/2}$ is positive for all real $a$.

Since the integrand has even parity $\phi(a)$ is odd. The two branches in (16) then appear as in Figure 1, with scaling of $a$ so that $a_+(0) = -a_-(0) = 1$. The branch $a_+(t)$ is shown in solid blue and the $a_-(t)$ branch as a red dashed curve. A smooth differentiable solution valid through the singularity now has odd parity (everywhere) - consistent with $a \sim 1/t$. The Friedmann equation 'predicts' a post singularity universe that is a mirror image of (our) pre-singularity universe though with $a \to -a$. For the Friedmann equation to remain valid the image must be a legal – dynamically feasible – copy of the original pre-singularity universe.

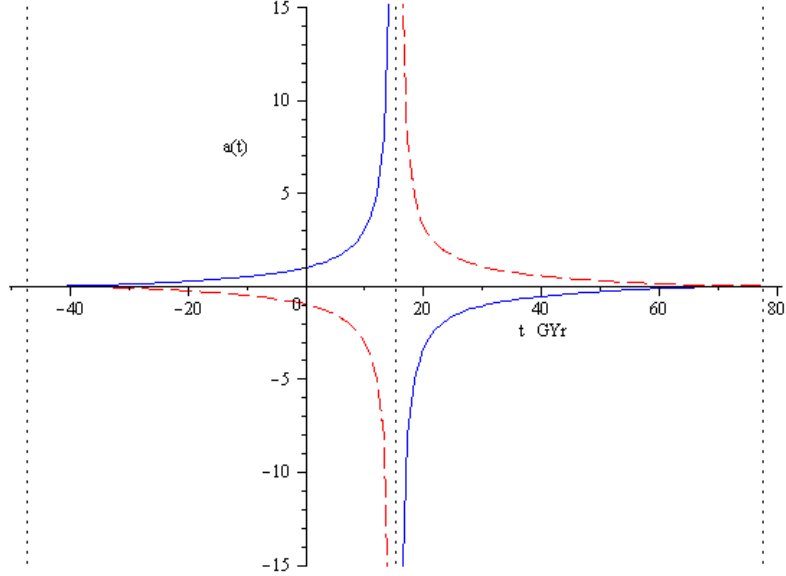

Figure 1. Plot of both branches of the scale factor on both sides of the conformal singularity; the solid curves are a single branch and the broken curves are a single branch. Here the origin of the time coordinate has been chosen so that $a(0) = 1$.

### 3.4 Singularities and Asymptotic Behavior

The time to the singularity found by integrating (9) numerically is

$$t(a=\infty) - t(a=1) = \frac{1}{H} \int_1^\infty \frac{da}{\sqrt{\Omega_{EM} + \Omega_m a + \Omega_\Lambda a^4}} = 1.12 / H = 15.2 \text{ Gyr} \qquad (18)$$

which is marked in Figure 1 by a dotted line. The conformal time elapsed since the Big Bang – the age - is

$$t(a=1) - t(a=0) = \frac{1}{H} \int_0^1 \frac{da}{\sqrt{\Omega_{EM} + \Omega_m a + \Omega_\Lambda a^4}} = 3.47 / H = 47.2 \text{ Gyr} . \qquad (19)$$

The conformal interval from Big Bang to future singularity is therefore 62.4 GYr, which we will call the duration. Focusing on the branch $a_+(t)$, the Big Bang was at time $t = -47.2$ GYr ago. The scale factor expands from 0 to positive infinity at 15.2 Gyr in the future. The scale factor then changes sign and proceeds to diminish in magnitude from negative infinity to 0, which it reaches at time 15.2 + 62.4 = 77.6 Gyr.

Near the future singularity the vacuum-dominated asymptotic behavior is that of a simple pole

$$a(t) \to (1 - kt)^{-1} . \qquad (20)$$

Therefore (20) with $k = 1/15.2$ Gyr is an approximation to the remaining evolution which ignores the non-asymptotic behavior. In the following we will exploit the fact that (9) is invariant under time translations to move the time of the future singularity to $t = 0$ with the result that the present time is *negative* 15.2 Gyr. The evolution near the singularity is then like $a \sim 1/t$ and is odd about the new origin. This has the advantage that if the transformation (6) is applied to this system (with this origin) then the conformal singularity occurs at the same time for both systems defined in (3) i.e. at $t = 0$. Further, the asymptotic behavior $a \sim 1/t$ corresponds in the system with $f(x) = a(t/x^2)/x^2$ to $f(x) \sim 1/t$ also, so both systems have the same asymptotic behavior. After all this, near the singularity we can then ignore the differences because the two forms are essentially the same. We should point out however that these are not the only forms of conformal factor that represent de Sitter spacetime. The curved-space Robertson-Walker spacetime for example has a different de Sitter asymptote (Lasenby, 2002; Lasenby, A. & Doran, C., 2005) which is not covered by the analysis here, and would be interesting to investigate in this context.

## 4. EM IN CONFORMAL SPACETIME

The EM action in curved spacetime is

$$I = -\int d^4x \sqrt{-g} \left( \frac{1}{4} F_{ab} F^{ab} + A_a j^a \right). \tag{21}$$

In the particular case of conformal spacetime it is useful to re-write this as

$$I = -\int d^4x \left( \frac{1}{4} F_{ab} F_{bd} \eta^{ac} \eta^{bd} + A_a \sqrt{-g} j^a \right). \tag{22}$$

The covariant divergence of the current must vanish

$$j^a_{;a} = 0 \Rightarrow \partial_a \left( \sqrt{-g} j^a \right) = 0. \tag{23}$$

Apart from this, the equations are the same as for Minkowski spacetime. Therefore it is convenient to define

$$\bar{j}^a = \sqrt{-g} j^a \tag{24}$$

whose ordinary divergence must now vanish. With this, the action (22) becomes

$$I = -\int d^4x \left( \frac{1}{4} F_{ab} F_{bd} \eta^{ac} \eta^{bd} + A_a \bar{j}_b \eta^{ab} \right) \tag{25}$$

and now the scale factor has been eliminated. It follows that variation of the covariant potentials in (25) must give the Maxwell equations as if in Minkowski spacetime:

$$\partial^2 A_a - \partial_a (\partial \circ A) = \bar{j}_a. \tag{26}$$

The Minkowski spacetime Lorenz gauge

$$\partial \circ A \equiv \eta^{ab} \partial_a A_b = 0 \tag{27}$$

then leads to

$$\partial^2 A_a = \bar{j}_a. \tag{28}$$

Appendix A gives the discrete inversion symmetries in Minkowski spacetime. It will be necessary for subsequent discussions to know how these are affected in going to conformal spacetime. Of particular interest is the fate of time reversal symmetry due to the presence of a scale factor that is odd about the conformal singularity. Since the scale factor is absent in the Lorenz gauge, the symmetric partner of $A^\mu(x)$ is $h^\mu_{\ \nu} A^\nu(-\tilde{x})$ just as in Minkowski spacetime, though this is demoted to a point symmetry rather than a local one.

Consider now the consequences of imposing instead the covariant gauge condition

$$A^\mu_{\ ;\mu} = \frac{1}{\sqrt{-g}} \partial_\mu \left( \sqrt{-g} A^\mu \right) = 0 \Rightarrow \partial_\mu \left( a^4 A^\mu \right) = 0 \Rightarrow 2\frac{\dot{a}}{a} \phi + \frac{\partial \phi}{\partial t} + \nabla \cdot \mathbf{A} = 0 \tag{29}$$

where the 1 + 3 potentials are components of a covariant vector:

$$\{A_\mu\} = (\phi, \mathbf{A}); \quad A^\mu = g^{\mu\nu} A_\nu \Rightarrow \{a^4 A^\mu\} = a^2 (\phi, \mathbf{A}). \tag{30}$$

Then (26) is replaced by

$$\partial^2 \phi + 2\frac{\dot{a}}{a}\frac{\partial \phi}{\partial t} + 2\left(\frac{\ddot{a}}{a} - \frac{\dot{a}^2}{a^2}\right)\phi = \bar{\rho}, \quad \partial^2 \mathbf{A} = \bar{\mathbf{j}} + 2\frac{\dot{a}}{a}\nabla\phi \tag{31}$$

where we used

$$\{\bar{j}_\mu\} = (\bar{\rho}, \bar{\mathbf{j}}). \tag{32}$$

Evidently the components of the current vector are not affected by this change and so transform under time reversals as they did before. It follows from the structure of (31) that the potentials are likewise unaffected and the symmetric partner of $A^\mu(x)$ remains $h^\mu_\nu A^\nu(-\tilde{x})$. Hence the column headed $A^\mu(x)$ in Table 1 remains as it was in Table A2. The fate of EM fields in conformal spacetime is discussed in more detail in (Ibison, 2010).

## 5. THE DIRAC EQUATION IN CONFORMAL SPACETIME

### 5.1 Discrete Symmetries
In going to conformal spacetime from Minkowski spacetime we need only consider the effects of space and time inversions, since charge and mass inversion are unchanged. Mass inversion however will be 'opposed' by any operation that inverts the sign of the conformal factor. Discrete symmetries broken by the Cosmological metric have been discussed by Tomaschitz (Tomaschitz, R., 1994).

### 5.2 Time Reversal
Here we return to (1), restoring the conformal factor in one of the forms (3), and re-examine the effects of the inversions. In both cases the Dirac equation is no longer time-translation invariant due to the cosmological evolution, though local time reversal symmetry remains approximate valid in our era, far from the conformal singularity. Consider for example the rest energy of the Dirac particle computed from (A26). This is now

$$\langle E \rangle = a(t) m \int d^3x \, \psi^\dagger \gamma^0 \psi \tag{33}$$

which is not in general time-independent. Using (20) (which has an origin $t = 0$, $a(0) = 1$ corresponding to 'now') far from the boundary the energy is monotonically increasing at a rate

$$\frac{d}{dt}\langle E \rangle \approx km \int d^3x \, \psi^\dagger \gamma^0 \psi \tag{34}$$

whereas the energy monotonically decreases for a time-reversed particle. This is just the local version of Cosmological red-shift seen from the perspective of the conformal coordinate system. Red-shift is associated in this coordinate system with an increase in rest mass, whilst the EM fields are unaffected by the expansion. Generally we think of this effect as observable only as a result of interactions over great distances. In principle though the effect could be probed locally according to (34), leading to a local determination of the cosmological arrow of time through a broken time reversal symmetry.

In the presence of cosmological expansion exact time reversal symmetry, if it exists, relates two objects either side of the conformal singularity, i.e. as a point symmetry. Both conformal factors in (3) change sign under time reversal but are otherwise unaffected by the inversions. This changes the sign of $mf(x)$ from what it was in the Minkowski case. That sign change can be accommodated in the Dirac equation by replacing $\psi$ with $\gamma^5\psi$, which change is reflected in the rows for $\mathcal{T}_+$ and $\mathcal{T}_-$ in Table 1 compared with Table A2.

### 5.3 Parity Inversion
Parity inversion has no effect on the conformal factor $f(x) = a(t)$. Space translation invariance appears to be absent in the case of $f(x) = a(t/x^2)/x^2$. We assume however a freedom to choose a coordinate system with the spatial origin centered at the particle of interest. To be more precise, we can exploit the translation invariance of the $f(x) = a(t)$ system and then transform using (6) to the system $f(x) = a(t/x^2)/x^2$, which is equivalent to having performed a space-time translation within a paraboloidal surface of simultaneity. From this perspective parity inversion remains a local symmetry in both coordinate systems, as reflected by the entries in Table 1.

**5.4 Summary**
The discrete symmetries of Minkowski spacetime are preserved in a cosmological conformal expansion, though with some modifications. The biggest change is in time-reversal symmetry, which ceases to become locally valid, but retains a point symmetry. The lack of a local time-symmetry manifests as a 'Cosmological arrow of time', associated in particular with recession of distant galaxies in the Hubble flow and with Cosmological red-shift.

|  | $i$ | $\langle E \rangle$ | e | m | x | $mf(x)$ | $A^\mu(x)$ | constraints on EM coupling for point symmetry to exist | constraints on EM coupling for local symmetry to exist | symmetric partner of $\psi(x)$ | Dirac representation |
|---|---|---|---|---|---|---|---|---|---|---|---|
| $\mathcal{P}_+$ | + | + | + | + | $\tilde{x}$ | + | $h^\mu_{\ \nu} A^\nu(\tilde{x})$ | $A^\mu(\tilde{x}) = A^\mu(x)$ | $A^\mu(x) \to A^\mu(t)$ | $\gamma^0 \psi(\tilde{x})$ | $\gamma^0 \psi(\tilde{x})$ |
| $\mathcal{P}_-$ | - | - |  |  |  |  |  |  |  | $\left(\gamma^0 C \gamma^5 \gamma^0 \psi(\tilde{x})\right)^*$ | $i\gamma^2 \gamma^5 \gamma^0 \psi^*(\tilde{x})$ |
| $\mathcal{T}_+$ | + | + | + | + | $-\tilde{x}$ | - | $h^\mu_{\ \nu} A^\nu(-\tilde{x})$ | $A^\mu(-\tilde{x}) = A^\mu(x)$ | no exact local symmetry | $C^{-1*} \psi^*(-\tilde{x})$ | $\gamma^2 \gamma^0 \psi^*(-\tilde{x})$ |
| $\mathcal{T}_-$ | - | - |  |  |  |  |  |  |  | $\gamma^0 \gamma^5 \psi(-\tilde{x})$ | $\gamma^0 \gamma^5 \psi(-\tilde{x})$ |
| $\mathcal{C}_+$ | + | + | - | + | x | + | $-A^\mu(x)$ | $A^\mu$ held fixed (not negated) | none | $\left(\gamma^0 C \psi(x)\right)^*$ | $-i\gamma^2 \psi^*(x)$ |
| $\mathcal{C}_-$ | - | - |  |  |  |  |  |  |  | $\gamma^{0*} C^* \gamma^{5*} \gamma^0 C \psi(x)$ | $-\gamma^5 \psi(x)$ |
| $\mathcal{M}_+$ | + | - | + | - | x | - | $A^\mu(x)$ | none | none | $-\gamma^5 \psi(x)$ | $-\gamma^5 \psi(x)$ |
| $\mathcal{M}_-$ | - | + |  |  |  |  |  |  |  | $\left(\gamma^0 C \psi(x)\right)^*$ | $-i\gamma^2 \psi^*(x)$ |

Table 1. Inversions in Conformal Spacetime with Scale Factor $f(x) \sim 1/t$.

## 6. THE DIRAC WAVEFUNCTION IN CONFORMAL SPACETIME

### 6.1 Tetrad Formulation of the Dirac Action
The Dirac equation in Minkowski spacetime is

$$\left(\gamma^\alpha (i\hbar \partial_\alpha - eA_\alpha) - m\right)\psi = 0 \tag{35}$$

where $\alpha$ is a Lorentz index. In curved spacetime the effects of gravitation can be accounted for with the replacement

$$\begin{aligned} \partial_\alpha &\to \partial_\mu + \Gamma_\mu(x) \\ \gamma^\alpha &\to \gamma^\mu(x) = V_\alpha^{\ \mu}(x) \gamma^\alpha \end{aligned} \tag{36}$$

where the argument $x$ for the gamma-matrix signifies a different object from the Minkowski spacetime matrix, $g^\mu(x) \neq g^\mu$ and where $V_\alpha^{\ \mu}$ is a tetrad satisfying

$$g_{\mu\nu}(x) = V^\alpha_{\ \mu}(x) V^\beta_{\ \nu}(x) \eta_{\alpha\beta}$$

and therefore

$$g^{\mu\nu}(x) = V^{\alpha\mu}(x) V^{\beta\nu}(x) \eta_{\alpha\beta} = V_\alpha^{\ \mu}(x) V_\beta^{\ \nu}(x) \eta^{\alpha\beta}$$

$\Gamma_\mu$ is the spin connection defined by

$$\sigma\left[\Gamma_\nu(x), \gamma^\mu(x)\right] = \partial_\nu \gamma^\mu(x) + \Gamma^\mu_{\nu\rho} \gamma^\rho(x) \tag{37}$$

where $\sigma$ is introduced to compare different published results (see historical note below). In conformal spacetimes

$$g_{\mu\nu}(x) = f^2(x)\eta_{\mu\nu}.$$

The simplest choice is

$$V^\alpha{}_\mu(x) = f(x)\delta^\alpha{}_\mu \Rightarrow V_\alpha{}^\mu(x) = \delta_\alpha{}^\mu / f(x) \Rightarrow \gamma^\mu(x) = \gamma^\mu / f(x). \tag{38}$$

Then the spin connection equation is

$$\sigma\left[\Gamma_\nu(x), \gamma^\mu\right] f(x) = -\gamma^\mu \partial_\nu f(x) + \Gamma^\mu_{\nu\rho} \gamma^\rho \tag{39}$$

where the affine connection for the conformal metric is

$$\Gamma^\mu_{\nu\rho} = \left(\delta^\mu_\rho \partial_\nu + \delta^\mu_\nu \partial_\rho - \eta_{\nu\rho}\partial^\mu\right) f(x). \tag{40}$$

Putting this into (39) gives

$$\sigma\left[\Gamma_\nu(x), \gamma^\mu\right] = \left(\delta^\mu_\nu \slashed\partial - \gamma_\nu \partial^\mu\right)\phi; \quad \phi = \log f(x) \tag{41}$$

whose solution is

$$\Gamma_\nu(x) = -\frac{\sigma}{2}\gamma_\nu \slashed\partial \phi \tag{42}$$

(plus an arbitrary function times a constant matrix). Bearing in mind (36), one has

$$\gamma^\alpha \partial_\alpha \psi = \slashed\partial \psi \to \frac{1}{f}\gamma^\mu\left(\partial_\mu - \frac{\sigma}{2}\gamma_\mu \slashed\partial \phi\right)\psi = \frac{1}{f}\left(\slashed\partial - 2\sigma \slashed\partial \phi\right)\psi$$

$$\Rightarrow \slashed\partial \psi \to f^{2\sigma-1}\slashed\partial\left(\frac{\psi}{f^{2\sigma}}\right) \tag{43}$$

and so the Dirac equation in conformal spacetime is

$$i\hbar f^{2\sigma-1}\slashed\partial\left(\frac{\psi}{f^{2\sigma}}\right) - \frac{e}{f}\slashed A \psi - m\psi = 0. \tag{44}$$

Defining the normalized wavefunction $\tilde\psi = \psi / f^{2\sigma}$, and making the dependencies explicit, this becomes

$$\left(\gamma^\alpha(i\hbar\partial_\alpha - eA_\alpha(x)) - f(x)m\right)\tilde\psi(x) = 0. \tag{45}$$

### 6.2 Current Conservation

Note that the ordinary divergence of the current vanishes

$$\tilde\jmath^\alpha = e\bar{\tilde\psi}\gamma^\alpha\tilde\psi; \quad \partial_\alpha \tilde\jmath^\alpha = 0 \tag{46}$$

and therefore the charge is conserved in Minkowski space. Of course one can revert at any time to the un-normalized wavefunction using $\psi = f^{2\sigma}\tilde\psi$, for which the current obeys

$$\bar\jmath^\mu = e\bar\psi\gamma^\mu\psi, \quad \jmath^\mu{}_{;\mu} = \frac{1}{\sqrt{-g}}\partial_\mu\left(\sqrt{-g}\bar\jmath^\mu\right) = f^{-4}\partial_\mu\left(f^4 \bar\jmath^\mu\right) = 0. \tag{47}$$

(The symbol $\bar\jmath$ is used for the invariant current for consistency with the notation in the discussion of EM.) This is consistent with $\bar\jmath^\mu = f^{4\sigma}\tilde\jmath^\mu$ and (46) only if $\sigma = -1$, which fixes the correct value of $\sigma$ in (39):

$$\Gamma_\nu(x) = \frac{1}{2}\gamma_\nu \slashed\partial \phi. \tag{48}$$

### 6.3 A note on related work

Barut with others (Barut, A. O. & Duru, I. H., 1987; Barut, A. O. & Singh, L. P., 1995) give the spin connection equation (39) with $\sigma = 1$. Their equation is employed (Huang, 2005; Parashar, D., 1991) and otherwise widely cited elsewhere. An explicit equation of the form (39) with the correct sign is given has been given by Kovalyov (Kovalyov, M. & Légaré, M., 1989). Others (Birrell, N. D. & Davies, P. C. W., 1982; Villalba, V. M. & Percoco, U., 1989) bypass (39) and use the closed form result for the spin connection derived by Weinberg (Weinberg, S., 1972). That form can be shown to be compatible with (39) only for $\sigma = 1$. The result (42) has been given recently (Finster, F. & Reintjes, M., 2009), though the notation used by those authors is quite different.

**6.4 Summary**
It follows from the above that the Dirac equation in conformal spacetime with scale factor $f(x)$ is the same as the Dirac equation in Minkowski spacetime with the replacement

$$m \to f(x)m \tag{49}$$

whilst treating the current as conserved as usual (i.e. in Minkowski spacetime). It follows that an effective Lagrangian for the Dirac wavefunction in conformal spacetime is

$$I = -\int d^4x \,\bar{\psi}(x)\left(\gamma^\alpha\left(i\hbar\partial_\alpha - eA_\alpha(x)\right) - f(x)m\right)\psi(x) \tag{50}$$

where the $\gamma^\alpha$ are ordinary (Minkowski) gamma matrices. (Note that the scaling here differs from (Birrell & Davies, 1982); here the wavefunction is normalized for current conservation in Minkowski spacetime, un-weighted by the conformal factor.)

Though much has been written about the Dirac equation in conformal spacetime, this general result seems not to have been noticed, though there have been solutions given for the de Sitter case (see below) which effectively employ (50). The omission is probably due to the practice of working in the traditional Robertson-Walker coordinates rather than in conformal coordinates, which, especially in the case of curved space, tends to obscure the possibility of a reduction to (49).

**7. WAVEFUNCTION IN DE SITTER SPACETIME**

**7.1 Feynman - Gell-Mann Method of Solution**
Here we employ (45) to solve for the particular case of de Sitter spacetime with $f(x) = a(t) = 1/(Ht)$, regarded here as the asymptotic limit of the vacuum-dominated $K = 0$ Robertson-Walker cosmology. Given the above finding, henceforth we drop the tilde on the wavefunction and presume to be working solely in Minkowski spacetime with a dynamic mass. Then the free-space Dirac wavefunction (45) obeys

$$\left(i\gamma^\mu\partial_\mu - \lambda/t\right)\psi = 0; \quad \lambda = \frac{m_0 c^2}{\hbar H} \tag{51}$$

$\lambda$ is a dimensionless number of order $10^{40}$. We apply the Feynman - Gell-Mann method and make the substitution

$$\psi(t,\mathbf{x}) = \left(i\gamma^\mu\partial_\mu + \lambda/t\right)\phi(t,\mathbf{x}). \tag{52}$$

Then $\phi$ satisfies

$$\left(i\gamma^\mu\partial_\mu - \lambda/t\right)\left(i\gamma^\mu\partial_\mu + \lambda/t\right)\phi = \left(\partial^2 + \lambda\left(\lambda + i\gamma^0\right)/t^2\right)\phi = 0. \tag{53}$$

Separating out the spatial dependence of a single Fourier mode:

$$\phi(t,\mathbf{x}) = e^{i\mathbf{k}\cdot\mathbf{x}}\phi(t;\mathbf{k}), \quad \psi(t,\mathbf{x}) = e^{i\mathbf{k}\cdot\mathbf{x}}\psi(t;\mathbf{k}) \tag{54}$$

the bi-spinor $\phi(t;\mathbf{k}) = \chi(z); \quad z = |\mathbf{k}|t$ satisfies

$$\left(\frac{d^2}{dz^2} + 1 + \frac{\lambda(\lambda + i\gamma^0)}{z^2}\right)\chi(z) = 0 \tag{55}$$

and now (52) gives

$$\psi(t,\mathbf{k}) = \left(i\gamma^0\frac{\partial}{\partial z} - \hat{\slashed{k}} + \frac{\lambda}{z}\right)\chi(z). \tag{56}$$

In the Dirac representation

$$\gamma^0 = diag(1,1,-1,-1), \quad \chi^T = (u_+, \ v_+, \ u_-, \ v_-) \tag{57}$$

where $(u_+, v_+)$ is a positive energy spinor and $(u_-, v_-)$ is a negative energy spinor, and

$$\psi(t,\mathbf{k}) = \left( \begin{pmatrix} \lambda/z + i\partial_z & 0 \\ 0 & \lambda/z - i\partial_z \end{pmatrix} - \hat{K} \right) \chi(z) . \tag{58}$$

(In this notation $u_+$ and $v_-$ are both spin up, $u_-$ and $v_+$ are both spin down.) With (57), (55) becomes

$$\left( \frac{d^2}{dz^2} + 1 + \frac{\lambda(\lambda \pm i)}{z^2} \right) u_\pm = 0 . \tag{59}$$

The solution can be written in terms of Bessel functions which we choose to write as Hankel functions

$$u_\pm = \sqrt{z} H^{(1)}_{i\lambda \mp 1/2}(z), \quad v_\pm = \sqrt{z} H^{(2)}_{i\lambda \mp 1/2}(z) \tag{60}$$

Write the solutions of (56) as

$$\chi^T(z) = \sqrt{z} \left( H^{(1)}_{i\lambda-1/2}(z), \ H^{(2)}_{i\lambda-1/2}(z), \ H^{(1)}_{i\lambda+1/2}(z), \ H^{(2)}_{i\lambda+1/2}(z) \right) \tag{61}$$

with implicit coefficients for the initial conditions. Using that (Abramowitz, M. & Stegun, I. A., 1965)

$$\frac{d}{dz} H^{(i)}_\nu(z) = \frac{\nu}{z} H^{(i)}_\nu(z) - H^{(i)}_{\nu+1}(z)$$
$$\Rightarrow \frac{d}{dz}\left( \sqrt{z} H^{(i)}_\nu(z) \right) = \frac{(\nu + 1/2)}{\sqrt{z}} H^{(i)}_\nu(z) - \sqrt{z} H^{(i)}_{\nu+1}(z) \tag{62}$$
$$\Rightarrow (\lambda/z + i\partial_z)\left( \sqrt{z} H^{(i)}_{i\lambda-1/2}(z) \right) = -i\sqrt{z} H^{(i)}_{i\lambda+1/2}(z)$$

and that

$$\frac{d}{dz} H^{(i)}_\nu(z) = -\frac{\nu}{z} H^{(i)}_\nu(z) + H^{(i)}_{\nu-1}(z)$$
$$\Rightarrow \frac{d}{dz}\left( \sqrt{z} H^{(i)}_\nu(z) \right) = \frac{(-\nu + 1/2)}{\sqrt{z}} H^{(i)}_\nu(z) + \sqrt{z} H^{(i)}_{\nu-1}(z) \tag{63}$$
$$\Rightarrow (\lambda/z - i\partial_z)\left( \sqrt{z} H^{(i)}_{i\lambda+1/2}(z) \right) = -i\sqrt{z} H^{(i)}_{i\lambda-1/2}(z)$$

then

$$\begin{pmatrix} \lambda/z + i\partial_z & 0 \\ 0 & \lambda/z - i\partial_z \end{pmatrix} \chi(z) = \left( i\gamma^0 \partial_z + \lambda/z \right) \chi(z) = -i\gamma^5 \chi(z) . \tag{64}$$

Putting this in (58) gives

$$\psi(t,\mathbf{k}) = -\left( \hat{K} + i\gamma^5 \right) \chi(z) . \tag{65}$$

### 7.2 Independent Solutions

We wish now to generalize (65) to give the four independent solutions. From (55) any linear combination of $\chi(z)$ that commutes with $\gamma^0$ will remain a solution of that equation. In order for the solution of the Dirac equation to retain the structure of (65) we would need also to have the linear combination commute with $\hat{K}$ and $\gamma^5$. These two constraints combined are that the linear combinations must commute with $\gamma^a$; $a \in \{0,1,2,3,5\}$. The possibilities include the identity and pairwise combinations of $\gamma^i \gamma^j$; $i, j \in \{1,2,3\}$. One possibility is the matrix $Q$

$$Q := \begin{pmatrix} a & c & 0 & 0 \\ d & b & 0 & 0 \\ 0 & 0 & a & c \\ 0 & 0 & d & b \end{pmatrix} = a(1 + i\gamma^1\gamma^2)/2 + b(1 - i\gamma^1\gamma^2)/2 + c(i\gamma^2 + \gamma^1)\gamma^3/2 + d(i\gamma^2 - \gamma^1)\gamma^3/2 , \tag{66}$$

hence

$$\psi(t,\mathbf{k}) = \left(i\gamma^0 \frac{\partial}{\partial z} - \hat{\mathbf{K}} + \frac{\lambda}{z}\right) Q\chi(z)$$
$$= Q\left(i\gamma^0 \frac{\partial}{\partial z} - \hat{\mathbf{K}} + \frac{\lambda}{z}\right) \chi(z) \tag{67}$$
$$= -Q\left(\hat{\mathbf{K}} + i\gamma^5\right) \chi(z)$$

is a general solution, having 4 independent complex degrees of freedom. In order to relate the wavefunction to its initial state it will be more useful to express the constants as a vector:

$$Q\chi(z) = \Lambda(z)q; \quad q^T = (a,b,c,d). \tag{68}$$

Re-arranging (66) one obtains

$$\Lambda(z) = \sqrt{z} \begin{pmatrix} H^{(1)}_{i\lambda-1/2}(z) & 0 & H^{(2)}_{i\lambda-1/2}(z) & 0 \\ 0 & H^{(2)}_{i\lambda-1/2}(z) & 0 & H^{(1)}_{i\lambda-1/2}(z) \\ H^{(1)}_{i\lambda+1/2}(z) & 0 & H^{(2)}_{i\lambda+1/2}(z) & 0 \\ 0 & H^{(2)}_{i\lambda+1/2}(z) & 0 & H^{(1)}_{i\lambda+1/2}(z) \end{pmatrix}. \tag{69}$$

With this, the solution (67) can be written

$$\psi(t,\mathbf{k}) = -\left(\hat{\mathbf{K}} + i\gamma^5\right) \Lambda(z) q. \tag{70}$$

**7.3 Initial Conditions**

Let the initial conditions be that $\psi(t_0, \mathbf{k})$ is given. Then $q$ can be found from

$$\psi(t_0,\mathbf{k}) = -\left(\hat{\mathbf{K}} + i\gamma^5\right) \Lambda(z_0) q. \tag{71}$$

Observing that

$$\left(\hat{\mathbf{K}} + i\gamma^5\right)^{-1} = -\frac{1}{2}\left(\hat{\mathbf{K}} + i\gamma^5\right) \tag{72}$$

then $q$ is given by

$$q = \frac{1}{2}\Lambda^{-1}(z_0)\left(\hat{\mathbf{K}} + i\gamma^5\right)\psi(t_0,\mathbf{k}). \tag{73}$$

Making use of the wronskian (Abramowitz & Stegun, 1965)

$$H^{(1)}_{\nu+1}(z)H^{(2)}_{\nu}(z) - H^{(1)}_{\nu}(z)H^{(2)}_{\nu+1}(z) = -\frac{4i}{\pi z}, \tag{74}$$

the inverse of (69) is found to be

$$\Lambda^{-1}(z) = \frac{i\pi}{4}\sqrt{z} \begin{pmatrix} -H^{(2)}_{i\lambda+1/2}(z) & 0 & H^{(2)}_{i\lambda-1/2}(z) & 0 \\ 0 & H^{(1)}_{i\lambda+1/2}(z) & 0 & -H^{(1)}_{i\lambda-1/2}(z) \\ H^{(1)}_{i\lambda+1/2}(z) & 0 & -H^{(1)}_{i\lambda-1/2}(z) & 0 \\ 0 & -H^{(2)}_{i\lambda+1/2}(z) & 0 & H^{(2)}_{i\lambda-1/2}(z) \end{pmatrix}. \tag{75}$$

With this, the wavefunction at arbitrary times can be expressed explicitly in terms of the initial state:

$$\psi(t,\mathbf{k}) = -\frac{1}{2}\left(\hat{\mathbf{K}} + i\gamma^5\right)\Lambda(z)\Lambda^{-1}(z_0)\left(\hat{\mathbf{K}} + i\gamma^5\right)\psi(t_0,\mathbf{k}). \tag{76}$$

Cotaescu and Crucean (Coaescu, I. I. & Crusean, C., 2008) claimed to be the first to give a solution of (51) involving Hankel functions of complex order.

## 8. EFFECTS OF THE CONFORMAL SINGULARITY

### 8.1 Asymptotic Behavior Far From The Singularity

The region far from the singularity is characterized by $z \gg 1$ where $z = |\mathbf{k}|t$. Given the conformal time to the boundary computed in (19) is $t_{cb} = 47.2$ Gyr, for a particle to be in this region at the present time requires its speed $|\mathbf{v}|$ satisfy

$$|\mathbf{k}|ct_{cb} \gg 1 \Rightarrow |\mathbf{v}| \gg \frac{\hbar}{mct_{cb}} = \frac{c}{\omega_c t_{cb}}. \tag{77}$$

In practice in means the speed with respect to the Cosmological Frame must satisfy $|\mathbf{v}| \gg 10^{-31}$ m/s. For all practical purposes therefore, all matter in the present era is in the 'far field' of the future boundary.

We suppose that the initial condition is given a long way from the singularity, i.e. where $|z_0| \gg 1$. This is not the usual way of doing things, but here the conformal boundary is in the future and the wavefunction there will be presumed determined by propagation from an initially known state much earlier. Actually, the asymptotic expansions for the Hankel functions are given for large *positive* argument, which in our case is far in the future on the other side of the conformal boundary. Rather than try to change things around, we will compute the behavior of the wavefunction as it approaches the boundary from above – i.e. going backwards in time – assuming that the initial conditions (*q*) is given further into the future, with the asymptotic behavior at the boundary to be determined. In that case the mass is negative and $\lambda$ in (51) is large and negative (and real).

The magnitude of the Hankel functions are independent of order in the limit of large magnitude, so all the terms in (75) have equal weight. Use (Abramowitz & Stegun, 1965)

$$\sqrt{z_0} H_\alpha^{(1)}(z_0) \to \sqrt{\frac{2}{\pi}} e^{i(z_0 - \alpha\pi/2 - \pi/4)}, \quad \sqrt{z_0} H_\alpha^{(2)}(z_0) \to \sqrt{\frac{2}{\pi}} e^{-i(z_0 - \alpha\pi/2 - \pi/4)}. \tag{78}$$

Specifically:

$$\sqrt{z_0} H_{i\lambda+1/2}^{(1)}(z_0) \to -i\sqrt{\frac{2}{\pi}} e^{\lambda\pi/2 + iz_0}, \quad \sqrt{z_0} H_{i\lambda+1/2}^{(2)}(z_0) \to i\sqrt{\frac{2}{\pi}} e^{-\lambda\pi/2 - iz_0}$$

$$\sqrt{z_0} H_{i\lambda-1/2}^{(1)}(z_0) \to \sqrt{\frac{2}{\pi}} e^{\lambda\pi/2 + iz_0}, \quad \sqrt{z_0} H_{i\lambda-1/2}^{(2)}(z_0) \to \sqrt{\frac{2}{\pi}} e^{-\lambda\pi/2 - iz_0} \tag{79}$$

so $\Lambda^{-1}(z_0)$ in tends to

$$\Lambda^{-1}(z_0) \to i\sqrt{\frac{\pi}{8}} \begin{pmatrix} e^{-\lambda\pi/2 - iz_0} & 0 & ie^{-\lambda\pi/2 - iz_0} & 0 \\ 0 & e^{\lambda\pi/2 + iz_0} & 0 & -ie^{\lambda\pi/2 + iz_0} \\ e^{\lambda\pi/2 + iz_0} & 0 & -ie^{\lambda\pi/2 + iz_0} & 0 \\ 0 & e^{-\lambda\pi/2 - iz_0} & 0 & ie^{-\lambda\pi/2 - iz_0} \end{pmatrix}. \tag{80}$$

### 8.2 Wavefunction Near the Conformal Singularity

Near the boundary $z$ is small (and positive here) whereupon the Hankel functions approach the limiting forms (Abramowitz & Stegun, 1965)

$$\sqrt{z} H_\alpha^{(1)}(z) \to -\frac{i}{\pi}\left(\frac{2}{z}\right)^{\alpha-1/2} \Gamma(\alpha), \quad \sqrt{z} H_\alpha^{(2)}(z) \to \frac{i}{\pi}\left(\frac{2}{z}\right)^{\alpha-1/2} \Gamma(\alpha); \quad \text{Re}(\alpha) > 0. \tag{81}$$

Specifically:

$$\sqrt{z} H_{i\lambda+1/2}^{(1)}(z) \to -\frac{i}{\pi}\left(\frac{2}{z}\right)^{i\lambda} \Gamma(i\lambda+1/2), \quad \sqrt{z} H_{i\lambda+1/2}^{(2)}(z) \to \frac{i}{\pi}\left(\frac{2}{z}\right)^{i\lambda} \Gamma(i\lambda+1/2). \tag{82}$$

But (81) is valid only for real $\alpha$, so we will need to express the $H_{i\lambda-1/2}^{(j)}(z)$ differently before applying those limits. Using that

$$H_{i\lambda-1/2}^{(j)}(z) = -H_{i\lambda+3/2}^{(j)}(z) + \frac{2i\lambda+1}{z} H_{i\lambda+1/2}^{(j)}(z) \tag{83}$$

then

$$\sqrt{z}H^{(1)}_{i\lambda-1/2}(z) = -\sqrt{z}H^{(1)}_{i\lambda+3/2}(z) + \frac{2i\lambda+1}{\sqrt{z}}H^{(1)}_{i\lambda+1/2}(z)$$
$$\rightarrow \frac{i}{\pi}\left(\frac{2}{z}\right)^{i\lambda+1}\Gamma(i\lambda+3/2) - \frac{2i\lambda+1}{z}\frac{i}{\pi}\left(\frac{2}{z}\right)^{i\lambda}\Gamma(i\lambda+1/2). \tag{84}$$
$$= \frac{i}{\pi}\left(\frac{2}{z}\right)^{i\lambda+1}\Gamma(i\lambda+1/2)$$

Similarly

$$\sqrt{z}H^{(2)}_{i\lambda-1/2}(z) \rightarrow -\frac{i}{\pi}\left(\frac{2}{z}\right)^{i\lambda+1}\Gamma(i\lambda+1/2). \tag{85}$$

Putting (85), (84) and (82) into (69) gives

$$\Lambda(z) \rightarrow \frac{i}{\pi}\left(\frac{2}{z}\right)^{i\lambda}\Gamma(i\lambda+1/2)\begin{pmatrix} 2/z & 0 & -2/z & 0 \\ 0 & -2/z & 0 & 2/z \\ -1 & 0 & 1 & 0 \\ 0 & 1 & 0 & -1 \end{pmatrix}. \tag{86}$$

Near the boundary therefore

$$\Lambda(z) \rightarrow \frac{i}{\pi}\left(\frac{2}{z}\right)^{i\lambda+1}\Gamma(i\lambda+1/2)\begin{pmatrix} 1 & 0 & -1 & 0 \\ 0 & -1 & 0 & 1 \\ 0 & 0 & 0 & 0 \\ 0 & 0 & 0 & 0 \end{pmatrix}. \tag{87}$$

Since the entries in this matrix now have equal weight with respect to $\lambda$, the entries in the matrix $\Lambda^{-1}(z_0)$ can be assessed accordingly. The region $z > 0$ (to which the expansions are restricted) corresponds to the post-singularity half-space, in which the rest mass is negative. Therefore $\lambda < 0$, $|\lambda| \gg 1$ and only the exponentials of the form $\exp(-\lambda\pi/2)$ survive. The inverse $\Lambda^{-1}(z_0)$ then simplifies to

$$\Lambda^{-1}(z_0) \rightarrow i\sqrt{\frac{\pi}{8}}e^{|\lambda|\pi/2}e^{-iz_0}\begin{pmatrix} 1 & 0 & i & 0 \\ 0 & 0 & 0 & 0 \\ 0 & 0 & 0 & 0 \\ 0 & 1 & 0 & i \end{pmatrix} \tag{88}$$

where we have used that $\exp(-\lambda\pi/2) = \exp(|\lambda|\pi/2)$ to make the large size of the factor explicit. Combining (88) and (87):

$$\Lambda(z)\Lambda^{-1}(z_0) \rightarrow -\frac{1}{z\sqrt{2\pi}}\left(\frac{2}{z}\right)^{i\lambda}e^{|\lambda|\pi/2}e^{-iz_0}\Gamma(i\lambda+1/2)\begin{pmatrix} 1 & 0 & i & 0 \\ 0 & 1 & 0 & i \\ 0 & 0 & 0 & 0 \\ 0 & 0 & 0 & 0 \end{pmatrix}. \tag{89}$$

The gamma function has asymptotic behavior (Abramowitz & Stegun, 1965)

$$|\Gamma(i\lambda+x)| \rightarrow \sqrt{2\pi}|y|^{x-1/2}e^{-\pi|\lambda|/2} \Rightarrow |\Gamma(i\lambda+1/2)| \rightarrow \sqrt{2\pi}e^{-\pi|\lambda|/2}, \tag{90}$$

so (89) can be written

$$\Lambda(z)\Lambda^{-1}(z_0) \rightarrow \frac{1}{z}e^{i\phi(z,z_0,\lambda)}\begin{pmatrix} 1 & 0 & i & 0 \\ 0 & 1 & 0 & i \\ 0 & 0 & 0 & 0 \\ 0 & 0 & 0 & 0 \end{pmatrix} = \frac{1}{2z}e^{i\phi(z,z_0,\lambda)}(1+\gamma^0)(1+i\gamma^5) \tag{91}$$

where $\phi$ is a (real) phase. Putting this into (76), the wavefunction tends to a limit which, in terms of its initial value far from the boundary, is

$$\psi(t,\mathbf{k}) \rightarrow -\frac{1}{4z}e^{i\phi(z,z_0,\lambda)}(\hat{K}+i\gamma^5)(1+\gamma^0)(1+i\gamma^5)(\hat{K}+i\gamma^5)\psi(t_0,\mathbf{k}). \tag{92}$$

Defining the projection $P(\mathbf{k})$

$$P(\mathbf{k}) = \frac{1}{2}(1-\gamma^0)(1+\hat{\mathbf{k}}) = \frac{1}{k}\begin{pmatrix} 0 & 0 & 0 & 0 \\ 0 & 0 & 0 & 0 \\ -k_z & -k_x + ik_y & k & 0 \\ -k_x - ik_y & k_z & 0 & k \end{pmatrix} \qquad (93)$$

this becomes

$$\psi(t,\mathbf{k}) \to \frac{1}{z}e^{i\phi(z,z_0,\lambda)}P(\mathbf{k})\psi(t_0,\mathbf{k}). \qquad (94)$$

### 8.3 On the Reduction to two Dirac Components
We see from (93) that although all four components may be defined (as non-zero) away from the boundary, they are rolled up into just two components by the time they reach the boundary; only the *negative* energy components of the Dirac wavefunction are non-zero there.

With appropriately chosen asymptotic behavior for the Hankel functions the same procedure can be applied to the pre-boundary wavefunction, wherein the rest mass and therefore λ are positive. In that case one finds that only *positive* energy states arrive at the boundary - whatever the values of the 4 components specified in the 'initial conditions' away from the boundary.

On 'our' side at $t = 0_-$ the negative energy components are zero at the singularity. On the other side at $t = 0_+$ the positive energy components are zero. As it approaches the singularity the phase oscillates rapidly, infinitely quickly so at the boundary. Moving away from the singularity towards the present era – i.e. backwards in time – the appearance of non-zero negative energy components is due to subsequent interactions in which the initially pure positive energy state becomes mixed, and all four components in the Dirac wavefunction bi-spinor become occupied.

Note that this behavior by itself is *not* a boundary condition on the wavefunction. The negative energy states decay to zero as a consequence of intrinsic properties of the Dirac equation. If a boundary condition *were* to be imposed, it could only be on the positive energy states at $t = 0_-$ - regarded as a future boundary condition, and on negative energy states at $t = 0_+$ - regarded as an historical boundary condition for development of the post-singularity universe.

The analysis above culminating in (92) gives that only the positive energy components of the wavefunction components are non-zero at $t = 0_-$. It is expected that this behavior carries over into the second quantized theory so that electrons and positrons arrive at the boundary with zero velocity relative to the Hubble frame.

### 8.4 Hubble Drag
The 4-current in the Dirac representation is

$$\{j^\mu\} = \psi^\dagger \gamma^0 \{\gamma^\mu\}\psi = \psi^\dagger \begin{pmatrix} 1 & 0 \\ 0 & -1 \end{pmatrix}\left(\begin{pmatrix} 1 & 0 \\ 0 & -1 \end{pmatrix}, \begin{pmatrix} 0 & \boldsymbol{\sigma} \\ -\boldsymbol{\sigma} & 0 \end{pmatrix}\right)\psi = \psi^\dagger \left(\begin{pmatrix} 1 & 0 \\ 0 & 1 \end{pmatrix}, \begin{pmatrix} 0 & \boldsymbol{\sigma} \\ \boldsymbol{\sigma} & 0 \end{pmatrix}\right)\psi. \qquad (95)$$

As a result of the loss of the two negative energy components the 3-current vanishes:

$$\rho = e(|u_+|^2 + |v_+|^2), \quad \mathbf{j} = \mathbf{0}. \qquad (96)$$

An effect of the expansion is a 'Hubble Drag' that acts to bring all Dirac matter to rest in the Hubble frame. The same outcome is predicted of classical matter by the action

$$I = -m\int dt\, a(t)\sqrt{1-\mathbf{v}^2(t)}; \quad \mathbf{v}(t) := d\mathbf{x}(t)/dt. \qquad (97)$$

Here $t$ is the conformal time and $a \sim 1/t$. The effect of Hubble drag on the current is shown in Figure 2. A fuller discussion of this compared with the more traditional classical action

$$I = -m\int \sqrt{dx^\mu dx^\nu g_{\mu\nu}(x)} = -m\int dt|a(t)|\sqrt{1-\mathbf{v}^2(t)} \qquad (98)$$

can be found in (Ibison, 2010).

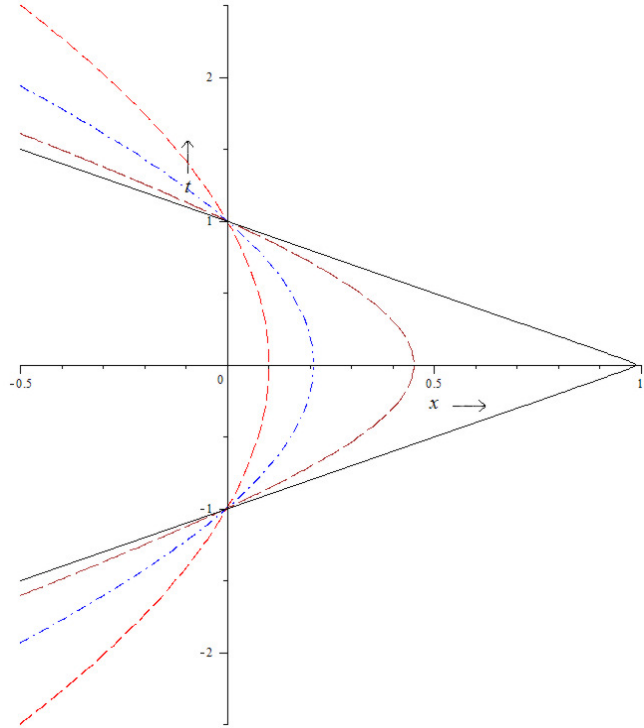

Figure 2. Geodesics of matter in the de Sitter limit of evolution for varying initial speeds specified far from the Conformal Singularity, which is at $t = 0$. The solid line is the geodesic for light speed as a limit approached from below. The time axis measures conformal time.

## 9. FUTURE COSMOLOGICAL BOUNDARY CONDITION

### 9.1 Relation between Pre and Post Singularity Universes
We will assume the laws of physics do not change either side of the singularity and in particular that Maxwell's equations and the single particle Dirac equation are obeyed everywhere, including at and through the conformal singularity. An assessment of the relationship between the pre and post singularity universes should take account of the following two important facts:
   i) The scale factor is anti-symmetric across the boundary, and
   ii) The pre-singularity wavefunction possesses symmetric partners across the boundary.

Because the Friedmann equation is symmetric about the boundary, the post-singularity evolution is a time-reversed version of the expansion on our side. It is hard to see how minor deviations can be tolerated at the particle level without damaging the symmetry of the scale factor, even though the latter signifies the development of only the most course-grain level of the Cosmological fluid. There is no scope in the Friedmann equation for minor deviations from perfect symmetry, so it is to be expected that this state of affairs be upheld consistently at every level, down to the individual particle wavefunctions. From consideration of the Friedmann equation alone therefore, one expects to find that the pre and post singularity Dirac wavefunctions are symmetric partners such that their contributions to the stress-energy at fixed **x** from both $t$ and $-t$ (relative to the singularity at $t = 0$) are identical.

We wish to make clear that there is no imperative from solving the Dirac equation alone that the pre and post singularity wavefunctions have any relationship. Indeed, the opposite seems true at first; the wavefunctions on either side of the singularity have no non-zero components in common and so are completely decoupled. Thus the input from the Friedmann equation that the two universes be related amounts to an extraneous constraint on the relationship between the two wavefunctions. It means that the two components that 'survive' on each side must somehow be related.

### 9.2 The 'No-Copy' Assumption
In the following we seek a picture that is consistent with both the symmetries of the Dirac equation in conformal spacetime and the Friedmann equation. A reasonable conclusion is that the post singularity universe is in some sense a copy of the pre-singularity universe. The most efficient explanation from an Ockham's Razor point of view is that there is no copy: the post-singularity universe and the post-singularity universe are physically the same universe. Then the geodesics mirrored in the conformal singularity in Figure 2 are to be taken as standing for the development of the ensemble as a whole, presumed therefore to be faithfully mirrored in detail at every level.

If we put the conformal singularity at $t = 0$, then the no copy constraint says that the universe at $t$ and $-t$, are the same, and therefore the state of affairs described physically on either side are related by an unobservable transformation. It does not have to mean that Cosmology is cyclic when parsed with monotonically increasing $t$ – it would only be so if the same condition were applied at the Big Bang. Given this, below we enumerate the possibilities consistent with the no-copy assumption.

### 9.3 Time-Reversed Image

The post-singularity wavefunction will be an identical copy of the pre-singularity wavefunction if the coordinate system is folded back upon itself in such a manner that the pre and post singularity coordinates $(t,\mathbf{x})$ and $(-t,\mathbf{x})$ are physically the same point. Then, given a pre-singularity solution $\psi(-t,\mathbf{x})$, $t > 0$ the 'no-copy' requirement is that there must exist a post-singularity solution $\psi(t,\mathbf{x}) = \eta\psi(-t,\mathbf{x})$, $t > 0$ where $\eta$ is a constant unobservable phase. We would prefer these be compatible with the point of view that the wavefunction crosses the boundary propagating according to business as usual from the perspective of the Dirac equation (51), and then folding the space along the conformal singularity to find the post-singularity evolution is identical with the pre-singularity evolution. There is a problem however: From the right-most column of Table 1 one sees that there is no symmetry corresponding to $\psi(t,\mathbf{x}) = \eta\psi(-t,\mathbf{x})$, $t > 0$ because time reversal introduces $\gamma^0$ which can be removed only with a parity inversion. Hence this possibility can be discounted.

### 9.4 Parity Inversion

Motivated by CPT invariance of QED, we seek a 'no-copy' interpretation of the post-singularity universe in which $(t,\mathbf{x})$ and $(-t,-\mathbf{x})$ are physically the same points. The 'no copy' requirement then becomes $\psi(t,\mathbf{x}) = \eta\psi(-t,-\mathbf{x})$, $t > 0$. Reading off from Table 1 the possibilities are

$$\eta = \mathcal{P}_\rho \mathcal{T}_\rho \mathcal{M}_- \text{ or } \mathcal{P}_\rho \mathcal{T}_\rho \mathcal{C}_+ \text{ or } \mathcal{P}_\rho \mathcal{T}_{\bar\rho} \mathcal{M}_+ \text{ or } \mathcal{P}_\rho \mathcal{T}_{\bar\rho} \mathcal{C}_-; \quad \rho \in [+,-] \quad, \bar\rho = -\rho. \tag{99}$$

The identical universe requires also that

$$A^\mu(-x) = A^\mu(x) \tag{100}$$

with no room for absorption of an arbitrary factor. Note this is a point symmetry, not a local one. With reference to Table 1, charge inversion is therefore discounted, leaving only

$$\eta = \mathcal{P}_\rho \mathcal{T}_\rho \mathcal{M}_- \text{ or } \mathcal{P}_\rho \mathcal{T}_{\bar\rho} \mathcal{M}_+ . \tag{101}$$

This is as far as we can proceed unambiguously. Possibly there is no physically meaningful distinction to be made between the alternatives, in which case attempting to isolate one is a waste of effort. Acknowledging this risk, we give the following argument in favor of a particular combination. We would like to ascribe the process of mass inversion solely with propagation of the Dirac function across the boundary, in accord with the effect of the scale factor in the Dirac equation. As we have said, the singularity swaps the positive and negative energy solutions, which is achieved by both $\mathcal{M}_+$ and $\mathcal{M}_-$ (proportional to $\gamma^5$ and $\gamma^2$ respectively). Despite the singular behavior of the Dirac wavefunction at the boundary we would prefer to regard this exchange as the limit of an analytical process, which does not seem possible if the operation is anti-linear. This singles out $\mathcal{M}_+$. Similar reasoning concerning the coordinate folding at the singularity singles out the real operations $\mathcal{P}_+ \mathcal{T}_-$. This reasoning isolates $\eta = \mathcal{P}_+ \mathcal{T}_- \mathcal{M}_+$ as the only viable candidate. There is a close connection with CPT invariance here. Unlike the latter, we have set ourselves the constraint that the wavefunction and the EM fields transform back to themselves with no room in the latter for a sign inversion which is the reason mass inversion was favored over charge conjugation.

As briefly discussed in an earlier work (Ibison, 2010) a parity inversion can be achieved in the de Sitter spacetime at the time of a pair of conformal singularities by identification of the two hypersurfaces of co-dimension 1 there (Calabi, E. & Marcus, L., 1962; Hawking, S. W. & Ellis, G. F. R., 1973). In our case, the parity inversion and the 'no-copy' assumption can be achieved through the identification of two 4-volumes. In the case of the first of (3) and putting the conformal singularity at $t = 0$, the two volumes are inversions of each other through the origin. If there are no topological implications of the Big Bang then the topology is

$$\mathbb{R}^4 / \sim, \quad x \sim -x \; \forall x \in \mathbb{R}^4 . \tag{102}$$

In practice the conformal time since the Big Bang is finite so this manifold is only partially occupied by our universe. A wavefunction in our half-space $t < 0$ is unaffected by this identification, though at $t = 0$ it requires $\psi(0,\mathbf{x}) = \psi(0,-\mathbf{x})$ for all $\mathbf{x}$. Not wanting to favor any particular location, this implies $\psi(0,\mathbf{x}) =$ constant - a future boundary condition on the development of the Dirac wavefunction in our half-space. In the second quantized theory it is expected that this condition will take a form that requires every particle be annihilated by its anti-particle at the boundary.

The effects of the boundary condition will not generally be noticeable far from boundary, and similar to that of EM fields as discussed in more detail elsewhere (Ibison, 2010). Since only Fourier modes having wavelengths approaching that of the distance to the boundary are appreciably affected, effects become noticeable locally only at very low speeds and perhaps also at very low accelerations. The possibility of a connection with Modified Newtonian Dynamics (MOND) should be mentioned because the threshold acceleration for the onset of MOND effects is around $a_0 \sim 10^{-10}$ m/s$^2 \sim cH$ as most recently reported (Swaters, R. A., Sanders, R. H., and McGaugh, S. S., 2010) and $H$ sets the length scale (specifically $c/H$) at which wavelengths one expects to notice departures from standard theories that do not include a future boundary. Discussion of the implications for entropy and the second law of thermodynamics are omitted here, though it is hard to see how these could escape unmodified.

## 10. SUMMARY

We examined in detail the effect of the conformal factor on the discrete symmetries usually present in the Dirac theory and analyzed the behavior of the Dirac wavefunction near the future conformal singularity. We found that only two of the 4 bi-spinor components survive to the boundary, one effect of which is to cause the geodesics to experience a drag towards the Hubble frame. We then presented arguments for the existence of a genuine boundary condition on the wavefunction at the future conformal singularity based upon the assumption there does not exist a redundant copy of the universe in the post-conformal singularity era. The future boundary condition affects the spectral decomposition of the wavefunction and this might be locally testable, i.e. at the present time.

## ACKNOWLEDGEMENTS

The author is very grateful to Anthony Lasenby for sharing his considerable expertise.

## APPENDIX A: DISCRETE SYMMETRIES IN MINKOWSKI SPACETIME

In order to catalog the effects of the conformal factor let us first review the discrete symmetries as they would be in Minkowski spacetime, i.e. with $f(x) = 1$ in (1).

$$\left(\gamma^\mu\left(-i\partial_\mu + eA_\mu(x)\right) + m\right)\psi(x) = 0 \ . \tag{A1}$$

We deal first with Lorentz scalars and vectors, and subsequently with the Dirac equation. This appendix is a review and compilation of material which can be found, for example, in (Itzykson, C. & Zuber, J.-B., 1985; Peskin, M. E. & Schroeder, D. V., 1995; Weinberg, S., 2005).

### A.1 Lorentz Quantities

Let $\mathcal{C}, \mathcal{P}, \mathcal{T}, \mathcal{M}$ denote charge parity time and mass inversions. By mass here we mean rest-mass, which is a Lorentz scalar. Mass and charge inversions are internal to the particle and not part of the Minkowski geometry and not, therefore, members of the Lorentz Group. Let $\mathcal{I}$ be any member of this set of discrete inversions; $\mathcal{I} \in [\mathcal{P}, \mathcal{T}, \mathcal{C}, \mathcal{M}]$, $\mathcal{I}^2 = 1$. Fundamentally their effects are just to change signs of the associated quantities as summarized in Table A1.

|  | $e$ | $m$ | $x$ |
|---|---|---|---|
| $\mathcal{P}$ | + | + | $\tilde{x}$ |
| $\mathcal{T}$ | + | + | $-\tilde{x}$ |
| $\mathcal{C}$ | - | + | $x$ |
| $\mathcal{M}$ | + | - | $x$ |

Table A1. Effects of charge, parity, time and mass inversions on $e$, $m$, and $x$.

Here, '−' denotes a sign inversion and

$$\tilde{x} = \{h^\mu{}_\nu x^\nu\}, \quad h^\mu{}_\nu := \begin{pmatrix} 1 & 0 & 0 & 0 \\ 0 & -1 & 0 & 0 \\ 0 & 0 & -1 & 0 \\ 0 & 0 & 0 & -1 \end{pmatrix}. \tag{A2}$$

The Lorentz scalar fields transform straightforwardly as

$$\mathcal{I}\left[f(x;e,m)\right] = f\left(\mathcal{I}^{-1}[x]; \mathcal{I}^{-1}[e], \mathcal{I}^{-1}[m]\right) = f\left(\mathcal{I}[x]; \mathcal{I}[e], \mathcal{I}[m]\right), \tag{A3}$$

the second step following because $\mathcal{I}^2 = 1$. To determine the effect on a Lorentz vector we could perhaps take as our template the derivative of a scalar field associated perhaps with a charge e and mass m: $f(x;e,m)$. It's transformation is easily determined from the substitutions $\mathbf{x} \to \mathbf{x}' = -\mathbf{x}$ so $\nabla \to -\nabla$ and $t \to t' = -t$ so $\partial/\partial t \to -\partial/\partial t$ [3]:

$$\begin{aligned}\mathcal{M}[\partial_\mu f(x;e,m)] &= \partial_\mu f(x;e,-m) \\ \mathcal{C}[\partial_\mu f(x;e,m)] &= \partial_\mu f(x;-e,m) \\ \mathcal{P}[\partial_\mu f(x;e,m)] &= h_\mu^{\ \nu}\partial_\nu f(\tilde{x};e,m) \\ \mathcal{T}[\partial_\mu f(x;e,m)] &= -h_\mu^{\ \nu}\partial_\nu f(-\tilde{x};e,m)\end{aligned} \qquad (A4)$$

However, it is traditional to regard the zeroth component of a Lorentz vector as something that keeps its sign even under time reversals. For example the retarded Liénard-Wiechert potentials of a source following a path $\mathbf{x}_s(t)$ are sometimes written

$$A^\mu(t,\mathbf{x}) = \frac{e(1, d\mathbf{x}_s(t')/dt')}{|\mathbf{x}-\mathbf{x}_s(t')| - (\mathbf{x}-\mathbf{x}_s(t')) \cdot d\mathbf{x}_s(t')/dt'}; \quad t - t' = |\mathbf{x}-\mathbf{x}_s(t')| \qquad (A5)$$

which does not have the transformation property $\mathcal{T}[A_\mu(x)] = -h_\mu^{\ \nu}A_\nu(-\tilde{x})$ that would be inferred from (A4). This is easiest to see in the limit of a static charge:

$$\phi(t,\mathbf{x}) = \frac{e}{|\mathbf{x}-\mathbf{x}_s|} \qquad (A6)$$

which shows no capacity for changing sign under time reversal. The effects of time-reversal are felt instead in the space part, due to the presence there of derivatives with respect to time. Bearing this in mind, and taking into account also that the dependence on charge is known explicitly: $A_\mu(x) = A_\mu(x;e) = \mathrm{sgn}(e)A_\mu(x;|e|)$, the inversions of the vector potential are

$$\begin{aligned}\mathcal{M}[A_\mu(x)] &= A_\mu(x) \\ \mathcal{C}[A_\mu(x)] &= -A_\mu(x) \\ \mathcal{P}[A_\mu(x)] &= h_\mu^{\ \nu}A_\nu(\tilde{x}) \\ \mathcal{T}[A_\mu(x)] &= h_\mu^{\ \nu}A_\nu(-\tilde{x})\end{aligned} \qquad (A7)$$

We will see below that $\mathcal{T}[A_\mu(x)] = h_\mu^{\ \nu}A_\nu(-\tilde{x})$ is consistent with the standard interpretation of the time reversal operator on the Dirac wavefunction and, consequently, on the 4-current.

---

[3] Another way of thinking about this is to regard the parity operation for example as converting the contra-variant to the covariant form and vice-versa:

$$\mathcal{P}[V^\mu(x)] = V_\mu(\tilde{x}) \text{ and } \mathcal{P}^2 = 1 \Rightarrow \mathcal{P}(V_\mu(x)) = V^\mu(\tilde{x}).$$

However this has the effect of scrambling the Lorentz indexes in a product of a vector with a gamma matrix (because the latter are treated as universal constants):

$$\mathcal{P}[\gamma^\mu V_\mu(x)] = \gamma^\mu \mathcal{P}[V_\mu(x)] = \gamma^\mu V^\mu(\tilde{x}).$$

Here we prefer to treat the transformation matrix as a mixed tensor as defined in (A2).

## A.2 Gamma Matrices and Slashed Vectors

In discussing whether or not these symmetries exist the gamma matrices are treated here as universal constants, and not subject to or affected by the inversions. It will be useful in the following to have in hand the inversion of the slashed vectors:

$$\begin{aligned} m\left[\gamma^\mu A_\mu(x)\right] &= \gamma^\mu A_\mu(x) \\ \mathcal{C}\left[\gamma^\mu A_\mu(x)\right] &= -\gamma^\mu A_\mu(x) \\ \mathcal{P}\left[\gamma^\mu A_\mu(x)\right] &= \gamma^\mu h_\mu{}^\nu A_\nu(\tilde{x}) = \gamma^{\mu\dagger} A_\mu(\tilde{x}) = \gamma^0 \gamma^\mu \gamma^0 A_\mu(\tilde{x}) \\ \mathcal{T}\left[\gamma^\mu A_\mu(x)\right] &= \gamma^0 \gamma^\mu \gamma^0 A_\mu(-\tilde{x}) \end{aligned} \tag{A8}$$

and

$$\begin{aligned} m\left[\gamma^\mu \partial_\mu\right] &= \gamma^\mu \partial_\mu \\ \mathcal{C}\left[\gamma^\mu \partial_\mu\right] &= \gamma^\mu \partial_\mu \\ \mathcal{P}\left[\gamma^\mu \partial_\mu\right] &= \gamma^0 \gamma^\mu \gamma^0 \partial_\mu \\ \mathcal{T}\left[\gamma^\mu \partial_\mu\right] &= -\gamma^0 \gamma^\mu \gamma^0 \partial_\mu \end{aligned} \tag{A9}$$

Inversion operations are symmetries of the Dirac equation if there exists an $\mathcal{I}[\psi(x)]$ compliant with (A2) for which

$$\mathcal{I}\left[\left(\gamma^\mu(-i\partial_\mu + eA_\mu(x)) + m\right)\psi(x)\right] = \mathcal{I}\left[\gamma^\mu(-i\partial_\mu + eA_\mu(x)) + m\right]\mathcal{I}\left[\psi(x)\right] = 0 \tag{A10}$$

where $\mathcal{I}[\psi(x)]$ can be written in terms of $\psi(x)$.

## A.3 Linear and Anti-Linear Inversion Operations on the Dirac Equation

Two possibilities are allowed for: $\mathcal{I}$ is linear and unitary, or anti-linear and anti-unitary. Let us distinguish between these two with subscripts as follows

$$\begin{aligned} \mathcal{I}_+[i] &= i, & \mathcal{I}_+[\psi(x)] &= U_+ \psi(\mathcal{I}_+^{-1}[x]) = U_+ \psi(\mathcal{I}_+[x]) \\ \mathcal{I}_-[i] &= -i, & \mathcal{I}_-[\psi(x)] &= U_- \psi^*(\mathcal{I}_-^{-1}[x]) = U_- \psi^*(\mathcal{I}_-[x]) \end{aligned} \tag{A11}$$

where $U_+, U_-$ are 4x4 matrices. The coordinates are insensitive to the distinction:

$$\mathcal{I}_+[x] = \mathcal{I}_-[x] = \mathcal{I}[x]. \tag{A12}$$

It will be useful to write $U_-$ in terms of $U_+$ so that just one of the two cases need be solved for. Applying (A11) and (A12) to (A10) gives

$$\left(\gamma^\mu \mathcal{I}_+\left[-i\partial_\mu + eA_\mu(x)\right] + \mathcal{I}[m]\right) U_+ \psi(\mathcal{I}[x]) = 0 \tag{A13}$$

and

$$\left(\gamma^\mu \mathcal{I}_-\left[-i\partial_\mu + eA_\mu(x)\right] + \mathcal{I}[m]\right) U_- \psi^*(\mathcal{I}[x]) = \left(\gamma^\mu \mathcal{I}_+\left[i\partial_\mu + eA_\mu(x)\right] + \mathcal{I}[m]\right) U_- \psi^*(\mathcal{I}[x]) = 0. \tag{A14}$$

Taking the conjugate of the latter:

$$\left(\gamma^{\mu*} \mathcal{I}_+\left[-i\partial_\mu + eA_\mu(x)\right] + \mathcal{I}[m]\right) U_-^* \psi(\mathcal{I}[x]) = 0. \tag{A15}$$

We introduce a charge conjugation matrix $C$ defined by the property

$$-\gamma^{\mu T} = C\gamma^{\mu} C^{-1} \tag{A16}$$

$C$ depends on the representation of the $\gamma^{\mu}$. We will need

$$\gamma^{\mu *} = -\gamma^0 C \gamma^{\mu} C^{-1} \gamma^0 = \gamma^0 C \gamma^5 \gamma^{\mu} \gamma^5 C^{-1} \gamma^0 \tag{A17}$$

with which (A15) can be written

$$\left(\gamma^{\mu} \mathcal{I}_+\left[-i\partial_{\mu} + eA_{\mu}(x)\right] + \mathcal{I}[m]\right)\gamma^5 C^{-1} \gamma^0 U_-^* \psi\left(\mathcal{I}[x]\right) = 0. \tag{A18}$$

Comparing with (A13) one infers that

$$U_+ \psi(\mathcal{I}[x]) = \kappa \gamma^5 C^{-1} \gamma^0 U_-^* \psi(\mathcal{I}[x]) \Rightarrow U_-^* = \eta^* \gamma^0 C \gamma^5 U_+ \Rightarrow U_- = \eta \gamma^{0*} C^* \gamma^{5*} U_+^* \tag{A19}$$

where $\eta$ and $\kappa$ are phase factors. In the Dirac representation

$$C = i\gamma^2 \gamma^0 = C^* \Rightarrow C^{-1} = i\gamma^0 \gamma^2 = -C \tag{A20}$$

and so (A19) is

$$U_- = \eta \gamma^0 C \gamma^5 U_+^* = i\eta \gamma^0 \gamma^2 \gamma^0 \gamma^5 U_+ = \lambda \gamma^2 \gamma^5 U_+^* \tag{A21}$$

where $\lambda$ is another phase factor. In the following we look at each inversion operation in detail, summarizing the results in Table A2.

**A.4 Parity Inversion**

From (A8) and (A9),

$$\mathcal{P}_+\left[\gamma^{\mu}\left(-i\partial_{\mu} + eA_{\mu}(x)\right) + m\right] = \gamma^0 \gamma^{\mu} \gamma^0 \left(-i\partial_{\mu} + eA_{\mu}(\tilde{x})\right) + m. \tag{A22}$$

We see that unless $A_{\mu}(\tilde{x}) = A_{\mu}(x)$, $\mathcal{P}_+$ is not a symmetry of the single particle Dirac equation. (An example where it is a symmetry is a central potential, for which $A_{\mu}(x) = (e/|\mathbf{x}|, \mathbf{0}) = A_{\mu}(\tilde{x})$). Hence the dynamics are not generally invariant under parity inversion, though the kinematics of the free particle may be. Note we are not investigating here whether or not parity inversion is a symmetry which, if applied to all of physics, including therefore the sources of $A_{\mu}$, leaves the dynamics unchanged. That is, we are not looking here for a *systematic symmetry* in the sense defined earlier. Instead, we are asking if there is a symmetric partner for a solution in a given field, which field is held constant whilst asking the question.

On the other hand, (A22) tells us that linear parity inversion is a *local symmetry* of the <u>free</u> Dirac particle equation with

$$\mathcal{P}_+\left[\psi(x)\right] = \eta \gamma^0 \psi(\tilde{x}). \tag{A23}$$

That is, for every solution $\psi(x)$ there exists another solution $\eta \gamma^0 \psi(\tilde{x})$. Here and in the following $\eta$ stands for a continuously re-definable phase factor which in each case we will choose so that the operator squares to 1. Hence in this case $\eta = 1$. We can read off from (A19) that the anti-linear version of parity inversion is

$$\mathcal{P}_-\left[\psi(x)\right] = \eta \left(\gamma^0 C \gamma^5 \gamma^0 \psi(\tilde{x})\right)^*. \tag{A24}$$

Using (A21) with (A23), in the Dirac representation this is

$$\mathcal{P}_-\left[\psi(x)\right] = \gamma^2 \gamma^5 \gamma^0 \psi^*(\tilde{x}) \tag{A25}$$

(with $\eta$ chosen so $\mathcal{P}_-^2 = 1$). We now examine the relationship between the energy of the original particle and its parity inverted partner. A stationary particle solving (2) varies in time as exp(-$i\gamma^0 mt$), so the original has expectation

$$\langle E \rangle = \int d^3x \psi^\dagger \left( i\frac{\partial}{\partial t} \right) \psi = m \int d^3x \psi^\dagger \gamma^0 \psi .  \tag{A26}$$

Replacing $\psi(x)$ with $\gamma^0 \psi(\tilde{x})$ changes the expectation to

$$\langle E \rangle = \int d^3x \psi^\dagger \gamma^{0\dagger} \left( i\frac{\partial}{\partial t} \right) \gamma^0 \psi = m \int d^3x \psi^\dagger \gamma^0 \psi . \tag{A27}$$

The components of the bi-spinor are chosen so that the mass and energy are positive in (A26). And whatever that choice, positivity of mass and energy are preserved in the parity-inverted case $\mathcal{P}_+$. The parity inverted solution corresponding to $\mathcal{P}_-$ varies as exp($i\gamma^0 mt$). Its stationary energy is therefore

$$\begin{aligned}
\langle E \rangle &= \int d^3x \left( \gamma^2 \gamma^5 \gamma^0 \psi^*(\tilde{x}) \right)^\dagger \left( i\frac{\partial}{\partial t} \right) \left( \gamma^2 \gamma^5 \gamma^0 \psi^*(\tilde{x}) \right) \\
&= -\int d^3x \psi^T(\tilde{x}) \gamma^0 \gamma^5 \gamma^2 \left( i\frac{\partial}{\partial t} \right) \left( \gamma^2 \gamma^5 \gamma^0 \psi^*(\tilde{x}) \right) \\
&= \int d^3x \psi^T(\tilde{x}) \left( i\frac{\partial}{\partial t} \right) \psi^*(\tilde{x}) \\
&= -\left( \int d^3x \psi^\dagger(\tilde{x}) \left( i\frac{\partial}{\partial t} \right) \psi(\tilde{x}) \right)^*
\end{aligned} \tag{A28}$$

This energy is negative if $\psi(x)$ has positive energy. In any case, its energy is negated with respect to that of $\psi(x)$.

### A.5 Time Reversal

From (A8) and (A9),

$$\mathcal{T}_+ \left[ \gamma^\mu \left( -i\partial_\mu + eA_\mu(x) \right) + m \right] = \gamma^{\mu\dagger} \left( i\partial_\mu + eA_\mu(-\tilde{x}) \right) + m . \tag{A29}$$

Taking the complex conjugate:

$$\begin{aligned}
&\left[ \gamma^{\mu\dagger} \left( i\partial_\mu + eA_\mu(-\tilde{x}) \right) + m \right]^* \left[ \mathcal{T}_+ [\psi(x)] \right]^* = 0 \\
&\Rightarrow \left( \gamma^{\mu T} \left( -i\partial_\mu + eA_\mu(-\tilde{x}) \right) + m \right) \left[ \mathcal{T}_+ [\psi(x)] \right]^* = 0
\end{aligned} \tag{A30}$$

Since $\gamma^{\mu T} = C \gamma^5 \gamma^\mu \gamma^5 C^{-1}$ then

$$\gamma^5 C^{-1} \left[ \mathcal{T}_+ [\psi(x)] \right]^* = \eta \psi(-\tilde{x}) \Rightarrow \mathcal{T}_+ [\psi(x)] = \eta^* C^{-1*} \gamma^{5*} \psi^*(-\tilde{x}) . \tag{A31}$$

It follows that $\mathcal{T}_+$ is not in general a symmetry of the coupled Dirac equation, unless perhaps $A_\mu(-\tilde{x}) = A_\mu(x)$, though it is a *local symmetry* of the free Dirac particle equation according to (A31). In the Dirac representation this is

$$\mathcal{T}_+ [\psi(x)] = \gamma^2 \gamma^5 \gamma^0 \psi^*(-\tilde{x}) \tag{A32}$$

(with $\eta$ chosen so $\mathcal{T}_+^2 = 1$). Applying (A21), the conjugate operator is

$$\mathcal{T}_- [\psi(x)] = \gamma^0 \psi(-\tilde{x}) . \tag{A33}$$

Following the steps in the section on parity inversion, the energy of a $\mathcal{T}_+[\psi(x)]$ solution (staying with the Dirac representation) is

$$\langle E \rangle = \int d^3x \left( \gamma^5 \gamma^0 \gamma^2 \psi^*(-\tilde{x}) \right)^\dagger \left( i \frac{\partial}{\partial t} \right) \gamma^5 \gamma^0 \gamma^2 \psi^*(-\tilde{x})$$
$$= i \int d^3x \, \psi^T(-\tilde{x}) \gamma^{2\dagger} \gamma^{0\dagger} \gamma^{5\dagger} \gamma^5 \gamma^0 \gamma^2 \frac{\partial \psi^*(-\tilde{x})}{\partial t}$$
$$= i \int d^3x \, \psi^T(-\tilde{x}) \frac{\partial \psi^*(-\tilde{x})}{\partial t} \qquad (A34)$$
$$= -i \int d^3x \, \psi^\dagger(-\tilde{x}) \frac{\partial \psi(-\tilde{x})}{\partial t}$$

If again the original solution varies as $e^{-i\gamma^0 mt}$ then that will have energy given by (A26) whilst $\psi(-\tilde{x}) \sim e^{i\gamma^0 mt}$ which in (A34) gives

$$\langle E \rangle = m \int d^3x \, \psi^\dagger \gamma^0 \psi \qquad (A35)$$

so that the energy of the $\mathcal{T}_+$ operation remains positive. For the $\mathcal{T}_-[\psi(x)] = \gamma^0 \psi(-\tilde{x})$ solution the energy is negative due to the change in sign of the exponent.

### A.6 Charge Conjugation

Charge makes its appearance in the Dirac equation only through the term $e\gamma^\mu A_\mu(x)$. If charge conjugation is applied to both the vector potential according to (A8) and the charge $e$ of the particle according to Table A1 represented by the wavefunction, then this term and therefore the Dirac equation as a whole is unaffected by inversion; charge conjugation is a systematic symmetry of QED.

More interesting is if, for a wavefunction solving Dirac's equation for a given potential (sourced by fixed charges), there exists another simply-related wavefunction associated with a oppositely-charged particle. In this case

$$\mathcal{C}[A_\mu(x)] = A_\mu(x), \quad \mathcal{C}[eA_\mu(x)] = -eA_\mu(x) \qquad (A36)$$

and then

$$\mathcal{C}_+\left[\gamma^\mu(-i\partial_\mu + eA_\mu(x)) + m\right] = -\gamma^\mu(i\partial_\mu + eA_\mu(x)) + m . \qquad (A37)$$

Charge conjugation is a symmetry if there exists a $\psi_c(x) = \mathcal{C}_+[\psi(x)]$ that solves

$$\left(-\gamma^\mu(i\partial_\mu + eA_\mu(x)) + m\right) \psi_c(x) = 0 . \qquad (A38)$$

Taking the complex conjugate:

$$\left(\gamma^{\mu*}(i\partial_\mu - eA_\mu(x)) + m\right) \psi_c^*(x) = 0 . \qquad (A39)$$

Using (A17) this is

$$\left(\gamma^0 C \gamma^\mu C^{-1} \gamma^0 (-i\partial_\mu + eA_\mu(x)) + m\right) \psi_c^*(x) = 0 . \qquad (A40)$$

This is (A1) provided

$$\psi(x) = C^{-1} \gamma^0 \psi_c^*(x) \Rightarrow \psi_c(x) = \mathcal{C}_+[\psi(x)] = \left(\gamma^0 C \psi(x)\right)^* . \qquad (A41)$$

In the Dirac representation:

$$C_+[\psi(x)] = (i\gamma^0\gamma^2\gamma^0\psi(x))^* = (-i\gamma^2\psi(x))^* = -i\gamma^{2*}\psi^*(x).$$ (A42)

From (A19)

$$C_-[\psi(x)] = \gamma^{0*}C^*\gamma^{5*}\gamma^0 C\psi(x)$$ (A43)

which, in the Dirac representation, is

$$C_-[\psi(x)] = -\gamma^0 C\gamma^5\gamma^0 C\psi(x) = -\gamma^0\gamma^2\gamma^0\gamma^5\gamma^0\gamma^2\gamma^0\psi(x) = -\gamma^5\psi(x).$$ (A44)

The energy of $C_+[\psi(x)]$ is

$$\langle E \rangle = \int d^3x \left(-i\gamma^2\psi^*(x)\right)^\dagger \left(i\frac{\partial}{\partial t}\right)\left(-i\gamma^2\psi^*(x)\right) = \int d^3x \psi^T(x)\gamma^{2\dagger}\gamma^2\left(i\frac{\partial}{\partial t}\right)\psi^*(x) = -\left(\int d^3x \psi^\dagger(x)\left(i\frac{\partial}{\partial t}\right)\psi(x)\right)^*$$ (A45)

which is the negative of the original state. The energy of $C_-[\psi(x)]$ is clearly positive.

**A.7 Mass Negation**

With the action as defined in Table A1, applying (A10) and using (A8) and (A9),

$$m_+\left[\gamma^\mu\left(-i\partial_\mu + eA_\mu(x)\right) + m\right] = \gamma^\mu\left(-i\partial_\mu + eA_\mu(x)\right) - m = -\left(\gamma^5\gamma^\mu\gamma^5\left(-i\partial_\mu + eA_\mu(x)\right) + m\right)$$ (A46)

one immediately has

$$m_+[\psi(x)] = -\gamma^5\psi(x) = C_-[\psi(x)].$$ (A47)

It follows that

$$m_-[\psi(x)] = C_+[\psi(x)] = (\gamma^0 C\psi(x))^*.$$ (A48)

Notice that $m_\pm = C_\mp$ is to be expected from a cursory consideration of their effect on the Dirac equation. The energies however are not the same as for the equivalent charge conjugations. If $\psi(x)$ is a solution with rest mass m, then $\gamma^5\psi(x)$ is a solution of the same equation but with rest mass m. I.E.

$$\psi(x;-m) = -\gamma^5\psi(x;m).$$ (A49)

The effects of all the inversions are summarized in Table A2.

|   | $i$ | $\langle E \rangle$ | e | m | x | $A^\mu(x)$ | constraints on EM coupling for point symmetry to exist | constraints on EM coupling for local symmetry to exist | symmetric partner of $\psi(x)$ | Dirac representation |
|---|---|---|---|---|---|---|---|---|---|---|
| $\mathcal{P}_+$ | + | + | + | + | $\tilde{x}$ | $h^\mu{}_\nu A^\nu(\tilde{x})$ | $A^\mu(\tilde{x}) = A^\mu(x)$ | $A^\mu(x) \to A^\mu(t)$ | $\gamma^0 \psi(\tilde{x})$ | $\gamma^0 \psi(\tilde{x})$ |
| $\mathcal{P}_-$ | - | - | | | | | | | $\left(\gamma^0 C \gamma^5 \gamma^0 \psi(\tilde{x})\right)^*$ | $\gamma^2 \gamma^5 \gamma^0 \psi^*(\tilde{x})$ |
| $\mathcal{T}_+$ | + | + | + | + | $-\tilde{x}$ | $h^\mu{}_\nu A^\nu(-\tilde{x})$ | $A^\mu(-\tilde{x}) = A^\mu(x)$ | $A^\mu(x) \to A^\mu(\mathbf{x})$ | $C^{-1*} \gamma^{5*} \psi^*(-\tilde{x})$ | $\gamma^2 \gamma^5 \gamma^0 \psi^*(-\tilde{x})$ |
| $\mathcal{T}_-$ | - | - | | | | | | | $\gamma^0 \psi(-\tilde{x})$ | $\gamma^0 \psi(-\tilde{x})$ |
| $\mathcal{C}_+$ | + | - | - | + | $x$ | $-A^\mu(x)$ | $A^\mu$ held fixed (not negated) | $A^\mu$ held fixed (not negated) | $\left(\gamma^0 C \psi(x)\right)^*$ | $-i\gamma^2 \psi^*(x)$ |
| $\mathcal{C}_-$ | - | + | | | | | | | $\gamma^{0*} C^* \gamma^{5*} \gamma^0 C \psi(x)$ | $-\gamma^5 \psi(x)$ |
| $\mathcal{M}_+$ | + | - | + | - | $x$ | $A^\mu(x)$ | none | none | $-\gamma^5 \psi(x)$ | $-\gamma^5 \psi(x)$ |
| $\mathcal{M}_-$ | - | + | | | | | | | $\left(\gamma^0 C \psi(x)\right)^*$ | $-i\gamma^2 \psi^*(x)$ |

Table A2. Inversions in Minkowski Spacetime

Notes: The constraints on the coupling in the case of parity and time inversion are always satisfied if $A^\mu = 0^\mu$ - i.e. there is no EM coupling. The symmetric partner under charge conjugation is the wavefunction of an oppositely charge particle (to that associated with the original solution $\psi(x)$) in the same field – i.e. without changing the sign of the charge of the sources of that field. Charge conjugation is a systematic symmetry of QED, in which case $\psi(x)$ is unaffected.